\documentclass[submission,copyright,creativecommons]{eptcs}
\providecommand{\event}{ACL2 2025} 

\providecommand{\thisvolume}[1]{This volume of EPTCS, Open Publishing Association}

\usepackage{iftex}
\usepackage{tikz}
\usetikzlibrary{shapes.geometric, arrows, decorations.pathreplacing, positioning, arrows.meta}

\tikzstyle{entity} = [ellipse, minimum width=3cm, minimum height=1cm, text centered, draw=black, fill=blue!30]
\tikzstyle{arrow} = [thick,->,>=stealth]

\usepackage{booktabs}
\usepackage{array}

\usepackage{listings}
\usepackage[scaled=0.85]{beramono}  
\lstnewenvironment{code}[1][]%
  {\noindent\minipage{\linewidth}\medskip 
   \lstset{basicstyle=\ttfamily\footnotesize,frame=single,#1}}
  {\endminipage}

\usepackage{float}    

\newfloat{codefloat}{h}{lop}
\floatname{codefloat}{Listing}

\ifpdf
  \usepackage{underscore}         
  \usepackage[english]{babel}
  \usepackage[T1]{fontenc}        
\else
  \usepackage{breakurl}           
\fi

\usepackage{algorithm}
\usepackage{algpseudocode}

\usepackage{amsmath} 
\usepackage{amsfonts} 
\usepackage{graphicx} 

\usepackage{hyperref}

\usepackage{subcaption}

\expandafter\def\expandafter\UrlBreaks\expandafter{\UrlBreaks
  \do\a\do\b\do\c\do\d\do\e\do\f\do\g\do\h\do\i\do\j%
  \do\k\do\l\do\m\do\n\do\o\do\p\do\q\do\r\do\s\do\t%
  \do\u\do\v\do\w\do\x\do\y\do\z\do\A\do\B\do\C\do\D%
  \do\E\do\F\do\G\do\H\do\I\do\J\do\K\do\L\do\M\do\N%
  \do\O\do\P\do\Q\do\R\do\S\do\T\do\U\do\V\do\W\do\X%
  \do\Y\do\Z}

\floatname{algorithm}{Algorithm}

\title{An Enumerative Embedding of the Python Type System in ACL2s}
\author{Samuel Xifaras \qquad Panagiotis Manolios \qquad Andrew T. Walter \qquad William Robertson
	\institute{Khoury College\\Northeastern University\\ Boston, Massachusetts, USA}
\email{\{xifaras.s,p.manolios,walter.a,w.robertson\}@northeastern.edu}
}
\def\titlerunning{An Enumerative Embedding of the Python Type System in ACL2s}
\def\authorrunning{S. Xifaras, P. Manolios, A. T. Walter, W. Robertson}

\begin{document}
\maketitle

\begin{abstract}
Python is a high-level interpreted language that has become an
industry standard in a wide variety of applications.
In this paper, we take a
first step towards using ACL2s to reason about Python code by developing an
embedding of a subset of the Python type system in ACL2s.
The subset of Python types we support includes many of the most
commonly used type annotations as well as user-defined types comprised
of supported types.  We provide ACL2s definitions of these types,
as well as \texttt{defdata} enumerators that are customized to provide code coverage
and identify errors in Python programs.  Using the ACL2s embedding, we
can generate instances of types that can then be used as inputs to
fuzz Python programs, which allows us to identify bugs in Python code
that are not detected by state-of-the-art Python type checkers. We evaluate our 
work against four open-source repositories, extracting their type information 
and generating inputs for fuzzing
functions with type signatures that are in the supported subset of
Python types. Note that we only use the type signatures of functions to
generate inputs and treat the bodies of functions as black boxes.
We measure code coverage, which ranges from
about 68\% to more than 80\%, and identify code patterns that hinder coverage
such as complex branch
conditions and  external file system dependencies.
We conclude with a discussion of the results and recommendations for future work. 

\end{abstract}

\section{Introduction}
Python is an industry-standard language that is used in software engineering disciplines ranging from web development to machine learning \cite{vandeusenpython2023}. 
Its versatility and ease of use have propelled it to its position as the second most popular programming language by usage on GitHub, behind only JavaScript \cite{noauthortop2022}. 
With this popularity has come many proposed improvements to the language, including type annotations which were introduced in Python Enhancement Proposal (PEP) 484 \cite{vanrossumpep2014}. 
This is a reasonable addition, as 
static typing in programming has been shown to have a host of benefits \cite{hanenbergempirical2014}. 
As these type annotations continue to be adopted by developers, they represent a rich source of data for application analysis. 

Since Python software has become ubiquitous, this paper is motivated by the need for robust software verification in Python.
Python code is typically tested with unit testing, which tests a property of the code in a single example scenario.
Although unit testing can be effective when done well, it has a reputation for being burdensome, and there is evidence that developers find it challenging--or are not motivated--to cover deeply nested code \cite{zhaitest2019}.
In this paper, we introduce an \textit{extensible}, \textit{enumerative} embedding of the Python type system that
can generate representative examples of simple and complex Python types, and evaluate its effectiveness 
by measuring code coverage on real-world open source repositories. 
The representative examples generated by this embedding can drive fuzzing and property-based testing (PBT), among other techniques. 

Importantly, the goal of this work is not to model the semantics of the type system. 
This is mostly because it is unclear what the semantics of the type system of Python are. 
Rak-amnouykit et al. \cite{rak-amnouykitpython2020}, for instance, find that two of the most popular 
Python type checkers, \textit{mypy} \cite{themypyprojectmypynodate} and \textit{pytype} \cite{noauthorpytypenodate},
implement and check different type systems. Nor is it the goal to model the semantics of Python, which are complex 
and subject to change in new Python versions. For a preview of the complex considerations involved in modeling 
the semantics of Python, including scoping and generators, see \cite{politzpython2013}. 
Instead, our goal is to facilitate fuzzing and PBT of Python programs with an embedding of the Python type system that 
supports extension and can be used to generate test cases. 

Our embedding is \textit{extensible} in that it can be extended with
new types. This is necessary as Python primitive types can be composed
in infinitely many ways with compound types such as lists, tuples, and
dictionaries. User-defined classes are also frequently used by Python
programmers, and we support embedding classes as record types, as long
as their field types are recursively representable in the
embedding. The embedding also supports extension with union types. How
the type system works and how it can be extended are discussed in
Section \ref{sect:type_registration}.
 
Our embedding is \textit{enumerative} because it leverages the enumerative 
data definition framework of ACL2s \cite{defdata}. We say a data type is 
enumerative when it is associated with an enumerator function that maps 
natural numbers to elements of the type. After the embedding is extended 
to include user-defined types, the enumerative property of the embedding 
allows examples of these types to be generated immediately. This is useful 
for test data generation in unit testing, producing large numbers of test 
cases for PBT, and producing seed inputs for 
fuzzing. We discuss this aspect of the embedding further in 
Section \ref{sect:custom_enumerator}.

We chose to implement this embedding in ACL2s because we 
view it as uniquely positioned to
support the goals of this work. Its logical foundation enables formal
reasoning and theorem proving for Python types and constructs, and
the ACL2s \texttt{defdata} framework~\cite{defdata} grants the
enumerative property of the embedding. Taken together, these
enable seamless integration with dynamic approaches, such as fuzzing
and PBT. Accordingly, the contributions of this paper are being used
in ongoing work related to fuzzing in Python.

In summary, we claim the following contributions in this paper:
\begin{itemize}
	\item \textbf{An extensible, enumerative embedding} of the Python type system in ACL2s. The code is open-sourced and available.\footnote{\href{https://github.com/acl2/acl2/tree/master/books/projects/python/embedding}{https://github.com/acl2/acl2/tree/master/books/projects/python/embedding}}
	\item \textbf{Custom enumerators} for Python primitive types. The enumerators are designed to produce representative examples of Python types. 
	\item \textbf{An evaluation} of code covered in four open source repositories with inputs generated by the embedding's enumerators. 
\end{itemize}

Note that conversion of the embedding's generated examples to Python objects is not specified in this paper. 
Details are given in Xifaras's master's thesis~\cite{xifarasleveraging2024}, 
and we also plan to publish a specification of the conversion process in a forthcoming software engineering paper. 
To motivate why this problem is interesting and nontrivial in the context of our ongoing work in fuzzing,
consider the fact that pickle~\cite{noauthorpicklenodate}, 
Python's built-in object serialization protocol, does not flatly encode the data. 
It instead encodes objects as instructions to a virtual machine that, when played back, 
reconstruct the original object~\cite{sultaniknever2021}. We have observed that this 
storage format is far from ideal for fuzzing and poses significant security concerns~\cite{xifarasleveraging2024}. 

We invite the community to extend the embedding and implement new features or program analyses on top of it. There are many features of the type system that are missing from our embedding, such as protocol types \cite{vanrossumpep5442014}, and new features are added with every update to the Python language. This leaves many opportunities for future work. 

The remainder of the paper is organized as follows. In Section \ref{sect:motivating_example}, we discuss some background information and a motivating example. Section \ref{sect:related_work} gives an overview of related work. Section \ref{sect:type_registration} describes how the model is embedded and how it can be extended. Section \ref{sect:custom_enumerator} contains definitions of custom enumerators that produce representative examples of Python primitive types. We provide examples of the usage of the embedding in Section \ref{sect:usage}, and we evaluate its performance on several open source repositories in Section \ref{sect:evaluation}. We discuss the results and future work in Section \ref{sect:future_work}, and conclude in Section \ref{sect:conclusion}.

\section{Background and Motivating Example}\label{sect:motivating_example}

As Python has seen increasing use across the software engineering industry, the need to collaborate in large-scale codebases has grown. This has led to the emergence of static typing in Python through the PEP-484 system of type hints \cite{vanrossumpep2014}. These hints are optional and not checked by the language implementation. \textit{Type checkers}, such as mypy \cite{themypyprojectmypynodate} and pytype \cite{noauthorpytypenodate}, have emerged to statically verify the correctness of these type hints. In fact, Python's type system semantics are heavily influenced by mypy's design decisions \cite{lehtosaloour2019,vanrossumpep2014}. Due, however, to the complex semantics of Python and the complexity of the type hints themselves \cite{rothpython2022}, these type checkers are neither sound nor complete \cite{rak-amnouykitpython2020}. When Python code is deployed in a production environment, false negatives are particularly dangerous because they could lead to unhandled crashes, resulting in reduced availability, customer dissatisfaction, and developer frustration. 

Consider, for instance, Listing \ref{listing:falsepositive1}. This
listing contains a function, \texttt{create\_decimal}, that takes an integer and a
floating-point number, converts them to their string representations,
concatenates them, then converts the result to a float and returns
it. This is a seemingly innocuous function. Its stated purpose is to construct a floating-point value out of an integer component
representing the digits to the left of a decimal point and a
floating-point value representing the digits to the right of a decimal
point. At the time of this writing, the latest versions of two Python
type checkers on Python 3.12, mypy (v1.14.1) and pytype (v2024.10.11),
do not report any errors in this definition. However, there are
several corner cases that this function does not account for that
trigger unhandled exceptions. For instance, consider when the floating-point 
argument takes a negative value, or a value of \texttt{inf}
(infinity), or \texttt{nan} (not-a-number). When an integer is
concatenated with the string \texttt{"nan"}, the result cannot be cast
back into a float, resulting in an exception. Property-based testing
with input data that covers these special-case floating-point values
can find this issue, highlighting the importance of both strong
testing discipline and representative test data.

\begin{codefloat}
\begin{code}[language=Python]
def create_decimal(whole_part: int, decimal_part: float) -> float:
    """Create a decimal number from the given whole part and decimal part"""
    return float(str(whole_part) + str(decimal_part).lstrip('0'))
\end{code}
\caption{Motivating example: A function that behaves well when \texttt{decimal\_part} is non-negative and not "\texttt{nan}", "\texttt{inf}", or "\texttt{-inf}".}\label{listing:falsepositive1}
\end{codefloat}

The embedding we present in this paper enables the discovery of this and other bugs in Python programs through the generation of representative test data. In the case of the function given in Listing \ref{listing:falsepositive1}, we find this bug because of the customized enumerative floating-point data type that accurately represents Python's \texttt{float} type. Our goal with this work is to augment type checking with dynamic execution and formal reasoning, thereby enabling Python developers to extract more value from their tooling and investment in type annotations. We hope this grants developers greater confidence in their code. 

\section{Related Work}\label{sect:related_work}
In this section, we cover various areas of the literature that are related 
to our work, and offer brief discussions about how our work fits in with each. 

\textit{Type annotations in Python.} The Python language has gone
through many iterations, in particular with its system of type
annotations (also referred to as type hints). Multiple PEP documents
have been published about Python's type system, such as PEP-484
\cite{vanrossumpep2014}, the specification of type hints in Python,
PEP-483 \cite{vanrossumpep4832014}, which specifies the theory
behind type hints, and PEP-544 \cite{vanrossumpep5442014} which
introduces structural subtyping into the language. Di Grazia and
Pradel \cite{digraziaevolution2022} perform a comprehensive study
of open source Python code and measure the state of type annotation
usage in the ecosystem. They find that there is an upward trend in
type annotation usage, but fewer than 10\% of code elements are
annotated. There are also distinct usage patterns among different
repositories, and they find that repositories with higher numbers of
contributors tend to utilize type annotations more. The upward 
trend in usage of Python type annotations is a positive signal for the 
potential adoption of our work. 

\textit{Type checking in Python.} We have chosen Python types as an 
aspect of Python to model in ACL2s because they are relatively
simple to model, we anticipate that type annotations' popularity will
continue to grow, and tooling for type checking in Python has several
issues. Third-party tooling for supporting the Python type system is
necessary because PEP-484 clearly states that it is not the intent of
the Python implementation to statically check types
\cite{vanrossumpep2014}. Static type checkers have therefore arisen to
offer compile-time guarantees of type safety. \textit{mypy}
\cite{themypyprojectmypynodate} and \textit{pytype}
\cite{noauthorpytypenodate} are notable examples. Rak-amnouykit et
al. \cite{rak-amnouykitpython2020} perform a study on the outputs of
these type checkers, and they find that mypy and pytype implement two
different type systems. They also find, in corroboration of the
results of Di Grazia and Pradel \cite{digraziaevolution2022}, that
statically detectable type errors often do not seem to inhibit
developers from committing code. This suggests that although type
annotations are seeing increased use, there is much work to do in
fixing type errors in Python code. Evidence that this is a practical
problem is provided by Rak-amnouykit et al. who find that these tools
emit false positives \cite{rak-amnouykitpython2020}. Finally, Roth
shows that Python type hints are Turing complete
\cite{rothpython2022}, indicating that type checking in Python is an undecidable
problem. No solution can be both sound and complete.

\begin{codefloat}
\begin{code}
(defun add-nonparametric-type (name defdata-ty)
  (setf (gethash name *type-table*)
        `(:name ,name 
          :kind "nonparametric" 
          :defdata-ty ,defdata-ty)))

(defun add-parametric-type (name defdata-ty-lambda)
  (setf (gethash name *type-table*)
        `(:name ,name 
          :kind "parametric" 
          :defdata-ty ,defdata-ty-lambda)))

(defun add-alias-type (name alias-of)
  (let ((name (string-downcase name))
        (alias-of (string-downcase alias-of)))
    (when (equal (gethash alias-of *alias-table*) name) 
          (error "It is illegal to set ~a as an alias for ~a because ~a is already an alias for ~a." name alias-of alias-of name))
    (setf (gethash name *alias-table*)
          alias-of)))
\end{code}
\caption{Definitions of \texttt{add-nonparametric-type}, \texttt{add-parametric-type}, and \texttt{add-alias-type}.}\label{listing:typeaddfns}
\end{codefloat}

\textit{Formal verification of Python.} Given Python's importance in
the modern software landscape, formal verification of its semantics is
an appealing academic pursuit. While the goal of the present work is 
not to formally verify Python programs, we consider these works related 
because they relate to the theme of tooling for Python software verification.
 Several attempts have been made to
formally specify or verify subsets of Python. Ranson formally
specifies an operational semantics of a heavily restricted subset of
Python 2 called \textit{IntegerPython}, which only has integers and
booleans as data types~\cite{ranson2008semantics}. \textit{IntegerPython} is implemented 
in the Isabelle/HOL proof assistant, and Ranson proves correctness of a Turing machine simulator
written in the language~\cite{ranson2008semantics}.  Politz et
al. implement a small-step operational semantics for Python, and they
contribute a translator from general Python programs to a "core
language" for which the semantics are modeled
\cite{politzpython2013}. They test their implementation on many test
cases from the CPython implementation. Smeding, in a master's thesis
regarded by Politz et al. as "sadly unheralded"
\cite{politzpython2013}, implements an executable semantics in literate Haskell and
tests it against 134 test cases
\cite{smeding2009executable}. Smeding's semantics are for Python 2.5,
however, which is a deprecated version of the language. Also in a
master's thesis, K{\"o}hl implements an operational, executable
semantics of Python using the $\mathbb{K}$ semantic framework
\cite{kohlexecutable2021}. K{\"o}hl's semantics are based on Python
3.7, which is also a deprecated version of the language, but closer to
the language's current state. To our knowledge, there are no works
that formally specify or verify any part of Python in ACL2/ACL2s,
rendering the present work the first to do so.

\begin{codefloat}
\begin{code}
(defun init-types ()
  (add-nonparametric-type "integer" 'acl2s::py-integer)
  (add-alias-type "int" "integer")
  (add-nonparametric-type "float" 'acl2s::py-float)
  (add-nonparametric-type "bool" 'acl2s::py-bool)
  (add-nonparametric-type "unicode-codepoint-string" 
                          'acl2s::unicode-codepoint-string)
  (add-alias-type "unicode" "unicode-codepoint-string")
  (add-alias-type "str" "unicode-codepoint-string")
  (add-alias-type "boolean" "bool")
  (add-parametric-type "list"
   (lambda (el-ty)
     (let ((elt-ty-sym (translate-type-to-defdata
                         (if (stringp el-ty) el-ty (alist-to-plist el-ty)))))
       `(acl2s::listof ,elt-ty-sym))))
  (add-parametric-type "dictionary"
   (lambda (key-ty val-ty)
     (let ((key-ty-sym (translate-type-to-defdata 
                         (if (stringp key-ty) key-ty (alist-to-plist key-ty))))
           (val-ty-sym (translate-type-to-defdata 
                         (if (stringp val-ty) val-ty (alist-to-plist val-ty)))))
       `(acl2s::map ,key-ty-sym ,val-ty-sym))))
  (add-parametric-type "fixedtuple"
   (lambda (&rest types)
     (let ((ty-syms (mapcar (lambda (ty)
                              (translate-type-to-defdata
                                (if (stringp ty) ty (alist-to-plist ty))))
                            types)))
       `(acl2s::list ,@ty-syms))))
  (add-nonparametric-type "nonetype" 'acl2s::py-none)
  (add-nonparametric-type "bytes" 'acl2s::py-bytes))
\end{code}
\caption{Initial setup of the type table.}\label{listing:types}
\end{codefloat}

\textit{Fuzzing in Python.} As previously mentioned, we are engaged in ongoing 
work on utilizing the enumerative embedding for fuzzing Python code. Fuzzing 
in Python is a nascent area of study as Python becomes increasingly widespread 
in industry. PyRTFuzz \cite{pyrtfuzz} is a recent paper that 
proposes an approach to fuzzing the Python interpreter, and claims 
several bug discoveries. Our ongoing work focuses on fuzzing 
arbitrary Python code, rather than the interpreter, and the 
ACL2s-based type example generation introduced in the present 
paper is used to create seed inputs for fuzzing. HypoFuzz~\cite{hatfield-doddshypofuzz2022}, 
maintained by Zac Hatfield-Dodds, is a Python library based on the 
PBT library Hypothesis \cite{maciverhypothesis2019} that uses 
advanced fuzzing techniques and long time budgets to find 
counterexamples to properties. In his master's thesis, Xifaras covers 
the embedding presented here in greater depth, as well as how it integrates with a larger 
fuzzing system~\cite{xifarasleveraging2024}. Experimental 
results on fuzzing in Python are also presented~\cite{xifarasleveraging2024}.

\textit{ACL2s.} The ACL2 Sedan (ACL2s)~\cite{dillinger-acl2-sedan,acl2s11} is an
extension of the ACL2 theorem prover\cite{acl2-car,acl2-acs,acl2-web}. On top of the capabilities of ACL2, ACL2s provides the
following:
1) a powerful type system via the \texttt{defdata} data definition
framework~\cite{defdata} and the \texttt{definec} and
\texttt{property} forms, which support typed definitions and
properties, 2) counterexample generation capability via the \texttt{cgen}
framework, which is based on the synergistic integration of theorem
proving, type reasoning and testing~\cite{cgen,harsh-fmcad,harsh-dissertation}, 3) a powerful termination analysis based on calling-context
graphs~\cite{ccg} and ordinals~\cite{ManoliosVroon03,ManoliosVroon04,MV05}, 4) an (optional) Eclipse IDE plugin~\cite{acl2s11}, and 5) the ACL2s systems programming framework
(ASPF)~\cite{acl2s-systems-programming} which enables the development
of tools in Common Lisp that use ACL2, ACL2s and Z3 as a
service~\cite{drew-z3,enumerative-data-types,invariant-discovery-game,ankit-mpmt,acl2-workshop-checker-paper}. 
Our work builds on ACL2s and uses its data definition framework to model Python types. 
Walter et al. also build on the enumerative data types in ACL2s, adding 
dependent types and showing how dependent type enumerators can 
generate a great breadth of examples of 802.11 Wi-Fi packets~\cite{enumerative-data-types}. 
These "enumerative data types with constraints" may become useful for 
this embedding in future work. 
We additionally leverage the aforementioned 
ACL2s systems programming framework in our work to enable integration 
with foreign function interfaces and other application libraries such as an HTTP server.

\section{Embedding Construction}\label{sect:type_registration}
To implement the embedding, we leverage ACL2s Systems Programming \cite{acl2s-systems-programming} to create an API in Common Lisp that makes calls which affect an underlying ACL2s theory (the "world"). We use ACL2s Systems Programming as a foundation because it simplifies interfacing with foreign systems. In our ongoing fuzzing work, for instance, we implement an HTTP server on top of the Common Lisp API that accepts requests to update the embedding with new types and get examples of embedded types. Xifaras describes this HTTP interface in \cite{xifarasleveraging2024}. At this stage of the implementation, the only ACL2s calls that are being made are \texttt{defdata} calls, which in turn make several calls to \texttt{defthm}. 

The embedding uses two data structures to track information about known types: a \textit{type table} and an \textit{alias table}. The type table is a hash table that maps type names to property lists (\textit{plists}) that contain metadata about the types. Lines 1-11 of Listing \ref{listing:typeaddfns} show the two functions that extend this table, \texttt{add-nonparametric-type} and \texttt{add-parametric-type}.

As shown in Listing \ref{listing:typeaddfns}, the values of the hash table are plists that have three keys, \texttt{:name}, \texttt{:kind}, and \texttt{:defdata-ty}. If the type's kind is non-parametric, \texttt{defdata-ty} takes the value of an S-expression containing the \texttt{defdata} definition syntax. See \cite{defdata} for a reference on this syntax. Otherwise, it takes a lambda which defines how to produce the \texttt{defdata} definition expression from the parameters of the type. 

Types may be known by different names, or a programmer may want to assign multiple names to the same underlying type. Type aliases enable this. Extension of the type alias table is done via the \texttt{add-alias-type} function, whose definition is also given in Listing \ref{listing:typeaddfns} (lines 13-19). This function takes two string values that represent the alias and the name to be aliased. They are both converted to lowercase (lines 14-15) to maintain case insensitivity. Lines 16-17 perform a simple cycle check, to ensure that the name to be aliased is not already an alias for the given alias. This check could be generalized to arbitrary-length cycles.

The model of the type system starts with a set of base types defined in Common Lisp as shown in Listing \ref{listing:types}. Note the use of \texttt{add-nonparametric-type}, \texttt{add-parametric-type}, and \texttt{add-alias-type} as defined in Listing \ref{listing:typeaddfns}. This set of initial types was mostly derived from the set of most commonly used types in annotations, as identified by the work of Di Grazia and Pradel \cite{digraziaevolution2022}. The \texttt{acl2s} symbols that are shown are associated with \texttt{defdata} definitions of the embedding.

The embedding can also be extended with complex types in Python. Any 
user-defined class that has field types that are recursively representable 
in the embedding can be admitted to the model, with the caveat that the 
embedding does not support self-referential or mutually recursive class 
definitions at this time. Admission of recursively representable types 
is implemented by an iterative type extraction procedure that continues 
until a maximum number of iterations has been reached or until a fixed point. 

This extraction process is given in Algorithm \ref{alg:type_registration}. 
Lines 1-4 set up the state variables. $S'$ is the final set of types, $S'_{prev}$ 
maintains the set from the previous iteration to check whether a fixed point 
has been reached. On line 3, $C$ is set to the domain of $A$, which is the 
set of extracted user-defined classes in the subject codebase. In this definition, 
$maxIters$, the maximum number of iterations to perform if no fixed point is 
reached, is set to 5. If the time budget allows, it is advisable to increase this value so that as many 
user-defined types can be registered as possible. The remainder of the lines in this algorithm 
define the main loop. Line 7 contains the check for whether a type can be 
registered under the current model. \textit{types} is a helper function that 
returns the types used in a function signature. \texttt{registerType} (line 8)
registers (i.e. admits) the type in the embedding. Refer to Xifaras for further
details on this extraction~\cite{xifarasleveraging2024}. 

\begin{algorithm}[htb]
	\small
	\caption{Type Registration}
	\label{alg:type_registration}
	\begin{algorithmic}[1]
		\Require Initial type set $S$; mapping of class name to set of attribute types $A$; mapping of class name to set of method signatures $M$
		\Ensure Final type set $S'$
		\State $S' \gets S$
		\State $S'_{prev} \gets S'$
		\State $C \gets \text{dom}(A)$
		\State $maxIters \gets 5$
		\For{$i \gets 1$ to $maxIters$}
			\For{each $c \in C$}
				\If{$A(c) \subseteq S'$ \textbf{and} $\left(\bigcup_{m \in M(c)} \text{types}(m)\right) \subseteq S'$}
					\State registerType($c$) 
					\State $S' \gets S' \cup \{ c \}$
				\EndIf
			\EndFor
			\If{$S' = S'_{prev}$} \Comment{check for fixed point}
				\State \textbf{break}
			\EndIf
			\State $S'_{prev} \gets S'$
		\EndFor\\
		\Return $S'$
	\end{algorithmic}
\end{algorithm}

Listing \ref{listing:classtest} contains examples of user-defined types 
that can be registered with the type extraction process. 
The second of the two classes, \texttt{TestClassB}, references the 
first, \texttt{TestClassA}. These class types are used in the definitions 
of two functions, \texttt{use\_a} and \texttt{use\_b}. The \texttt{defdata} 
calls used to embed this type information are given in 
Listing~\ref{listing:defdatacalls}. Note that the listed \texttt{defdata} 
calls fully represent all types that appear in this example. This implies 
that \texttt{use\_a} and \texttt{use\_b} are extractable as 
"appropriate functions," as defined in Section \ref{sect:evaluation}.

\begin{codefloat}
\begin{code}
from typing import List, Tuple

class TestClassA:
    def __init__(self, a: float, b: List[int]) -> None:
        self.a = a
        self.b = b

class TestClassB:
    def __init__(self, a: int, b: TestClassA) -> None:
        self.a = a
        self.b = b

def use_a(a: TestClassA) -> Tuple[float, List[int]]:
    return (a.a, a.b)

def use_b(b: TestClassB) -> Tuple[int, TestClassA]:
    return (b.a, b.b)

a_inst = TestClassA(3.5, [1, 2, 3])
b_inst = TestClassB(4, a_inst)

use_a(a_inst)
use_b(b_inst)
\end{code}
\caption{Example of Python class definitions and functions that use them.}\label{listing:classtest}
\end{codefloat}

\begin{codefloat}
\begin{code}
(DEFDATA TY1039 PY-INTEGER)  ;; int
(DEFDATA TY1041 PY-FLOAT)  ;; float
(DEFDATA TY1043 (LISTOF TY1039) DO-NOT-ALIAS T)  ;; List[int]

;; class TestClassA[a: float, b: List[int]]
(DEFDATA ACL2S::CLASSTEST.TESTCLASSA
         (DEFDATA::RECORD (A . ACL2S::TY1041) (B . ACL2S::TY1043))

;; class TestClassB[a: int, b: TestClassA]
(DEFDATA ACL2S::CLASSTEST.TESTCLASSB
         (DEFDATA::RECORD (A . ACL2S::TY1039)
          (B . ACL2S::CLASSTEST.TESTCLASSA))

;; Tuple[float, List[int]]
(DEFDATA TY1096 (LIST TY1041 TY1043))

;; Tuple[int, TestClassB]
(DEFDATA TY1100 (LIST TY1039 CLASSTEST.TESTCLASSB))
\end{code}
\caption{\texttt{defdata} calls issued by type extraction procedure when analyzing Listing \ref{listing:classtest}.}\label{listing:defdatacalls}
\end{codefloat}

\subsection{Note on Embedding Philosophy}

Note that we have chosen to adopt a conservative philosophy when embedding classes in Python. The runtime type system of Python has duck typing semantics, and Python supports runtime operations that add and remove arbitrary attributes of objects while preserving the Python \texttt{isinstance} relation, which checks that an object is an instance of a given class. This means that, at runtime, an instance of class \texttt{Bar} that behaves exactly like an instance of class \texttt{Foo} (if it walks like a duck, talks like a duck...) can be passed off as an instance of \texttt{Foo}, and an instance of \texttt{Foo} can be mutated to look like an instance of \texttt{Baz}, at which point code that operates on instances of \texttt{Foo} no longer recognize it as an object of type \texttt{Foo}, except by \texttt{isinstance}. Attempting to generate examples that capture the broadness of duck typing semantics may be "representative" of what is possible in Python, but this may lead to error reports that would be easily rejected by a user as false positives, since an instance of \texttt{Foo} that looks and acts like an instance of \texttt{Baz}, for instance, would never be created by their code. This is what we mean by "conservative." We take the type annotations and corresponding class definitions at face value, in the same way that mypy does \cite{lehtosalomypy2024}.

\section{Custom Enumerators}\label{sect:custom_enumerator}

In ACL2s, data types are \textit{enumerative}. This means that each type is associated with an enumerator function that maps the natural numbers to examples of the type \cite{defdata}. This is useful in the context of fuzzing and property-based testing because if one can define a type of data in ACL2s, one immediately has access to examples of it. We say that each data type in ACL2s has an enumerator \textit{attached} to it, and the attached enumerator can be changed programmatically.

In our ongoing work on fuzzing in Python, we found that the default ACL2s \texttt{defdata} enumerators for certain primitive types do not produce a wide variety of examples. These default enumerators are intended to produce examples that would be readable by a student in case one causes their code to fail \cite{harsh-fmcad}. We are not concerned with readability of the examples, so we instead defined custom enumerators that produce a much wider range of values suitable for fuzzing. The definitions of these custom enumerators are given in this section.

\subsection{Integers}

To test code that works with integers, we have created a custom enumerator for integers that generates small-magnitude and very large-magnitude integer values of positive and negative sign. The Python integer type has arbitrary precision, but Python code often interfaces with native code which uses machine integers that may be 8, 16, 32, or 64 bits. Feedback from running code annotated with Python integers can help the programmer narrow down the integer type that their code actually expects. For example, a function that makes a call to a native routine in the popular library \textit{numpy} \cite{harris2020array} may be annotated as taking a Python integer, but any integer that does not fit within 64 bits may cause unexpected behavior because the underlying native code expects a \texttt{numpy.int64} value. 
The custom enumerator generates integers from several cases with probabilities given in Table \ref{tab:integerenumcases}. For convenience of notation, where $l, i, h \in \mathbb{Z}$, let $P^{+}_{2}(l, h) := \{2^i \mid l \leq i \leq h\}$, $P^{-}_{2}(l, h) := \{-2^i \mid l \leq i \leq h\}$, and $P^{\pm}_{2}(l, h) := P^{+}_{2}(l, h) \cup P^{-}_{2}(l, h)$.

\renewcommand{\arraystretch}{1.5} 

\begin{table}[htb]
\small
\centering
\begin{tabular}{lc >{\centering\arraybackslash} m{10cm}}
\toprule
\textbf{Description} & \textbf{\%} & \textbf{Set} \\
\midrule
Sum of powers of two & 85 & $\{ a + b \mid a \in P^{\pm}_{2}(0, 64) \wedge b \in P^{\pm}_{2}(0, 16) \}$ \\ 
65-bit integers & 6 & $UnionAll(\{ \{a, -a\} \mid 2 \leq a \leq 2^{65} \})$ \\ 
Powers of 2, with off by one & 6 & $UnionAll(\{\{a, a-1, a+1\} \mid a \in P^{\pm}_{2}(0, 65)\})$ \\ 
One & 1 & $\{1\}$ \\ 
Zero & 1 & $\{0\}$ \\ 
Negative one & 1 & $\{-1\}$ \\ 
\bottomrule
\end{tabular}
\caption{Custom integer enumerator cases}\label{tab:integerenumcases}
\end{table}

As an example of how one may define a custom enumerator the integer enumerator source code is given in Listing \ref{listing:intenumerator}. Note the correspondence between this definition and Table \ref{tab:integerenumcases}. The probability distribution is given on line 5, in the expression \texttt{'(85 6 6 1 1 1)}. Lines 6-21 define the cases of the enumerator. In the first case, two values are generated then added. In the second case, three 32 bit integers are generated, and then a helper function, \texttt{make-nat-upto-2-expt-65}, is called to produce a 65-bit integer. In the third case, a power of two is generated, and an offset of either -1, 0, or 1 is selected by generating a random number between zero and two and subtracting one from it (lines 17-18).  The remaining three cases are trivial.

\begin{codefloat}
\begin{code}
(defun python-int-enum/acc (n seed)  
  (declare (xargs :mode :program))
  (declare (ignore n) (type (unsigned-byte 31) seed))
  (b* (((mv choice seed)
        (defdata::weighted-switch-nat '(85 6 6 1 1 1) seed)))
    (case choice
      (0 (b* (((mv val-64 seed) (signed-power-of-two-enum-seed 0 64 seed))
              ((mv val-16 seed) (signed-power-of-two-enum-seed 0 16 seed)))
           (mv (+ val-64 val-16) seed)))
      (1 (b* (((mv r1 seed) (defdata::genrandom-seed (1- (expt 2 31)) seed))
              ((mv r2 seed) (defdata::genrandom-seed (1- (expt 2 31)) seed))
              ((mv r3 seed) (defdata::genrandom-seed (1- (expt 2 31)) seed))
              (v (make-nat-upto-2-expt-65 r1 r2 r3))
              ((mv sign seed) (random-bool seed)))
           (mv (* (if sign 1 -1) (1+ v)) seed)))
      (2 (b* (((mv pow-2 seed) (signed-power-of-two-enum-seed 1 65 seed))
              ((mv constant+1 seed) (switch-nat-safe-seed 3 seed)))
           (mv (+ pow-2 (1- constant+1)) seed)))
      (3 (mv -1 seed))
      (4 (mv 0 seed))
      (t (mv 1 seed)))))

(defun python-int-enum (n)  
  (declare (xargs :mode :program))
  (b* (((mv x &) (python-int-enum/acc 0 n)))
    x))
\end{code}
\caption{The integer enumerator.}
\label{listing:intenumerator}
\end{codefloat}

\subsection{Strings}
Python supports Unicode strings, so our enumerator generates many different varieties of Unicode strings. 
The probabilities are broken down as specified in Table \ref{tab:stringenumcases}. Note the use of $C_{n}(S)$ notation. 
We let the notation $C_{n}(S)$ denote the set of strings of length at most $n$ composed of characters in $S$. 
Formally, $C_{n}(S) := \{c_0 \dots c_i \dots c_n \mid c_0 \in S \wedge \cdots \wedge c_i \in S \wedge \cdots \wedge c_n \in S\}$. 
The set ASCII denotes the set of ASCII characters. The Emoji set denotes the set of Emoji Unicode characters. 
The set Gr denotes the set of Greek letter characters. The set MathSym denotes the set of mathematical symbol characters. 
The set LtnDiac denotes the set of latin letters with diacritic marks (such as {\"a} and {\'a}). 
The set CmpEmoji denotes the set of compound emojis, which are emoji characters that span two or more codepoints.

Note that there is a distinction between string \textit{literals} and the \texttt{str} class in Python. 
There are several string literals in Python that have different semantics. The standard string literal, 
denoted with quotes (\texttt{""}), produces an instance of the \texttt{str} class. The \texttt{str} class, 
according to Python's documentation, represents a sequence of Unicode codepoints \cite{pythonsoftwarefoundationbuilt-types2025}. 
Our enumerator produces sequences of Unicode codepoints, satisfying this definition. 
Raw strings, denoted \texttt{r""}, also produce \texttt{str} instances, but escape sequences are ignored. 
"F-strings," short for format strings, are denoted \texttt{f""}. These string literals support \texttt{printf}-like 
interpolation of variables into strings, and they also produce \texttt{str} instances. Byte string literals, 
despite having similar syntax to the aforementioned literals (\texttt{b""}), produce \texttt{bytes} instances. 
The \texttt{bytes} type represents a sequence of 8-bit integers, and its representation in the embedding 
is defined in the \texttt{defdata} framework as shown in Listing \ref{pybytesdef}.

\begin{codefloat}
\begin{code}
(defnatrange u8 (expt 2 8)) ;; alternatively, (defdata u8 (range integer (0 <= _ < (expt 2 8))))
(defdata py-bytes (listof u8) :do-not-alias t)
\end{code}
\caption{The \texttt{defdata} of Python's \texttt{bytes} type.}\label{pybytesdef}
\end{codefloat}

\begin{table}[htb]
\small
\centering
\begin{tabular}{lc >{\centering\arraybackslash} m{10cm}}
\toprule
\textbf{Description} & \textbf{\%} & \textbf{Set} \\
\midrule
ASCII strings & 50 & $C_{10^4}(\text{ASCII})$ \\ 
Emoji strings & 2 & $C_{10^4}(\text{Emoji})$ \\ 
Greek-letter strings & 2 & $C_{10^4}(\text{Gr})$ \\ 
Mathematical symbols & 2 & $C_{10^4}(\text{MathSym})$ \\ 
Latin diacritics & 2 & $C_{10^4}(\text{LtnDiac})$ \\ 
Compound emojis & 2 & $C_{10^4}(\text{CmpEmoji})$ \\ 
Mixed strings & 40 & $C_{10^4}(\text{ASCII} \cup \text{Emoji} \cup \text{Gr} \cup \text{MathSym} \cup \text{LtnDiac} \cup \text{CmpEmoji})$ \\ 
\bottomrule
\end{tabular}
\caption{Custom string enumerator cases.}\label{tab:stringenumcases}
\end{table}

\subsection{Floats}

Python, like many other programming languages, has a \texttt{float} type, which represents a 64-bit double-precision IEEE floating-point number. Although ACL2s does not have a built-in \texttt{float} type, it supports arbitrary precision rational numbers. We observe that all values a floating-point number can take are rational numbers, except for the special values of \texttt{-inf}, \texttt{inf}, and \texttt{nan}. Therefore, we represent the Python floating-point type as a union between the ACL2s rationals and ACL2s representations of the special floating-point values.

In order to define a floating-point type that is representative of Python's float type, we define a custom enumerator that produces rational numbers in predefined interesting categories, as well as the aforementioned special case values. We use the same $P^{+}_2$/$P^{-}_2$/$P^{\pm}_2$ notation from the previous integer enumerator definition. There is bias towards generating "edge case" values that may be likely to trigger interesting behavior. Table \ref{tab:floatenumcases} contains the cases of the enumerator. Note that the terms "normal" and "subnormal" are used. A "normal" floating-point number is one that has no leading zeros in its mantissa. A "subnormal" floating-point number is one that has one or more leading zeros in its mantissa. 

\begin{table}[htb]
\small
\centering
\begin{tabular}{m{5cm}c >{\centering\arraybackslash} m{8cm}}
\toprule
\textbf{Description}  & \textbf{\%} & \textbf{Set} \\
\midrule
Rational numbers &  76 &
$\left\{ 
\begin{aligned}
    n / k \mid n, k & \text{ generated using} \\
                    & \text{the integer enumerator}
\end{aligned}
\right\}$ \\ 
Powers of 2 with small-magnitude exponents &  5 &
$UnionAll\left( \{\{a, a - 1, a + 1\} \mid \, a \in P^{\pm}_{2}(-64, 64)\} \right)$ \\ 
Powers of 2 with large-magnitude exponents & 5 &
$UnionAll\left(
\begin{aligned}
    \{\{a, a - 1, a + 1\} \mid & \, a \in P^{\pm}_{2}(65, 1024) \cup \\
                               & \, P^{\pm}_{2}(-1024, -65)\}
\end{aligned}
\right)$ \\ 
Min and max normal 32-bit floats & 3 &
$\left\{
\begin{aligned}
    & 2^{-126}, 2^{-126} + 1, 2^{-126} - 1, \\
    & 2^{127}(2^{-23} - 2), 2^{127}(2^{-23} - 2) - 1, \\
    & 2^{127}(2^{-23} - 2) + 1
\end{aligned}
\right\}$ \\ 
Min and max normal 64-bit floats, with off by one & 3 &
$\left\{
\begin{aligned}
    & 2^{-1022}, 2^{-1022} - 1, 2^{-1022} + 1, \\
    & 2^{1023}(2^{-52} - 2), 2^{1023}(2^{-52} - 2) - 1, \\
    & 2^{1023}(2^{-52} - 2) + 1
\end{aligned}
\right\}$ \\ 
Max integer representable as a 32 or 64-bit float &  2 &
$\{ 2^{24}, -2^{24}, 2^{53}, -2^{53} \}$ \\ 
Min and max subnormal 32 and 64-bit floats & 2 &
$\left\{
\begin{aligned}
    & 2^{-149}, -2^{-149}, 2^{-126}(1 - 2^{-23}), \\
    & -2^{-126}(1 - 2^{-23}), 2^{-1074}, -2^{-1074}, \\
    & 2^{-1022}(1 - 2^{-52}), -2^{-1022}(1 - 2^{-52})
\end{aligned}
\right\}$ \\ 
Not-a-number & 1 & $\{\texttt{nan}\}$ \\ 
Positive infinity & 1 & $\{\texttt{inf}\}$ \\ 
Negative infinity & 1 & $\{\texttt{-inf}\}$ \\ 
Negative zero & 1 & $\{-0\}$ \\
\bottomrule
\end{tabular}
\caption{Custom floating-point number enumerator cases.}\label{tab:floatenumcases}
\end{table}

\section{Usage}\label{sect:usage}
The functionality of this embedding is exposed through a Common Lisp API. In this section, 
we cover usage examples of several of these API calls. Available functions include a setter 
for the random seed, the \texttt{add-parametric-type} 
and \texttt{add-nonparametric-type} functions for extending the type system, and the 
\texttt{generate-examples} function for retrieving examples.

\begin{codefloat}
\begin{code} 
(include-book "top")                   ;; load the embedding's ACL2s book
:q                                     ;; quit into raw lisp
(load "api.lsp")                       ;; Load the backend module, which contains the API
(in-package :acl2s-python-types)       ;; "acl2s-python-types" is the name of the API package

(defvar *enum-random-state*)                ;; create variable to hold random state
(setf *enum-random-state* (make-cl-seed 1)) ;; Set seed

;; Generate 100 examples of floats
(generate-examples "float" 100 *enum-random-state*)

;; Register a union between integers, floats, and strings called intfloatstr
(register-union "intfloatstr" '("int" "float" "str"))

;; Generate 100 examples of intfloatstr
(generate-examples "intfloatstr" 100 *enum-random-state*)
\end{code}
\caption{Example of API usage.}\label{listing:usageexample}
\end{codefloat}

Before any API calls can be used, the environment must be initialized properly. 
Lines 1-7 of Listing~\ref{listing:usageexample} contain setup code that is 
needed to import the ACL2s book, exit into the Common Lisp environment, 
load the Common Lisp API module (\texttt{api.lsp}), then initialize the random state. 
In the code listing, the working directory is assumed to be the root of the
embedding implementation's source code. 

Now that the environment is initialized, examples of types can be generated. 
Line 10 of Listing \ref{listing:usageexample} contains a call to \texttt{generate-examples} 
that generates 100 examples of Python floating-point numbers. 
If you were to run this code, it would produce an S-expression containing rational 
numbers and occasionally data structures that encode special float values in 
Python such as \texttt{nan}. These values can be deserialized into Python values, 
but the procedure for this is beyond the scope of this paper. 

Union types can be admitted to the embedding via the \texttt{register-union} API. 
Line 13 of Listing~\ref{listing:usageexample} contains an example of embedding
a union of Python's integer, floating-point, and string types. 100 examples of 
this union type are then generated in the subsequent expression (line 16). 
The equivalent Python type annotations for this union are \texttt{int | float | str} 
and \texttt{Union[int, float, str]}.

\section{Evaluation}\label{sect:evaluation}

With this embedding, our hope is to pave the way to formal reasoning about Python's types and to enable fuzzing and property-based testing of Python code. For our work to be suitable for these use cases, the examples of types that the embedding generates must be representative of values that actual Python code interacting with those types would expect. To verify that our embedding satisfies this criterion, we perform an evaluation of code coverage in four open source repositories: \texttt{mypy}~\cite{noauthormypy2025}, \texttt{mindsdb}~\cite{noauthormindsdb2025}, \texttt{black}~\cite{noauthorblack2025}, and \texttt{manticore}~\cite{noauthormanticore2023}. These repositories were chosen because they were noted as having a high type annotation density in the type annotation study of Di Grazia and Pradel \cite{digraziaevolution2022}. Our focal evaluation goal is to verify that we cover code that we expect to cover, given that we only have knowledge of the types of function signatures. Importantly, we expect \textit{not} to cover code that has external file system dependencies, or code that has complex conditionals that block execution paths. 

\subsection{Setup}

To prepare the repositories for fuzzing using the enumerative data types 
in our embedding, we extract type information from 
each codebase, and iteratively extend the embedding to embed as 
much type information in the codebase as possible with respect to the current 
limitations of our implementation. Recall that this procedure is given in 
Algorithm \ref{alg:type_registration} and explained in Section \ref{sect:type_registration}. 
Further details of this information extraction are given in Xifaras's previous work~\cite{xifarasleveraging2024}. 

\begin{algorithm}[htb]
	\small
	\caption{Function Signature Extraction}
	\label{alg:function_extraction}
	\begin{algorithmic}[1]
		\Require Type set $S$; set of function signatures in codebase $F$
		\Ensure Set of fuzzable functions $G$
		\State $G \gets \emptyset$
		\For{each $f$ in $F$}
			\If{$\text{types}(f) \subseteq S$}
				\State $G \gets G \cup \{f\}$
			\EndIf
		\EndFor\\
		\Return $G$
	\end{algorithmic}
\end{algorithm}

After type registration is completed, \textit{function signature extraction} can take place. The definition for this procedure is given in Algorithm \ref{alg:function_extraction}. It iterates over all functions in the target  codebase, and adds functions for which all types in the signature are present in the set of recognized types $S$. The helper function \textit{types} is again used to extract this set from each signature.

The output of the function signature extraction step is a set of \textit{appropriate functions}. In summary, a function is appropriate if its signature is fully annotated and every type that appears in the signature is embedded in ACL2s. These functions are then fuzzed in the manner described in the following subsection.

\subsection{Experiment Design}

After performing the previously described setup, we perform a small fuzz testing experiment on the set of appropriate functions with the intent of gathering code coverage information. We use the \textit{coverage.py}~\cite{batcheldercoveragepy2023} 
library to measure coverage. By default, this library measures line coverage (although it does not count whitespace lines, and it counts statements that wrap onto multiple lines as a single line).

We perform five independent trials, following evaluation guidance from Klees et al. \cite{kleesevaluating2018}. In each trial, we fuzz each appropriate function using a stream of examples generated by the embedding's enumerators for 440 seconds. During fuzzing, the input-output samples are collected and stored, and "replayed" after fuzzing is complete to obtain code coverage.

\subsection{Results}
\begin{figure}[htb]
	\centering
	
\includegraphics[width=4in]{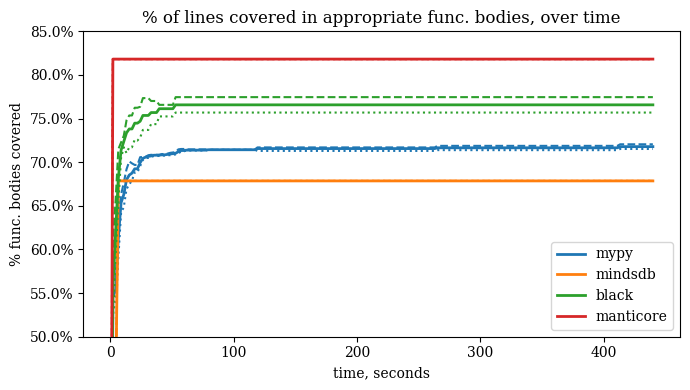}
	\caption{Results of coverage evaluation.}\label{fig:results}
\end{figure}

To provide insight into the coverage profile as the fuzz testing was taking place, we present 
coverage over time for the four repositories, averaged across the five independent trials. 
Figure \ref{fig:results} contains the coverage results. Code coverage for a fuzzing trial 
in a repository is measured as the percentage of covered statements in the union of the 
sets of statements in the bodies of that repository's appropriate functions. 
The solid lines presented in Figure \ref{fig:results} represent the average of this coverage. 
The dashed and dotted lines above and below each solid indicate 95\% confidence 
intervals for coverage. Note that the confidence interval lines are still being rendered for 
\texttt{mindsdb} and \texttt{manticore}, but they visually overlap with the solid average line.

Coverage is generally obtained very quickly. Note also that \texttt{mypy} and \texttt{black} 
seem to have longer "knees" on their curves than \texttt{mindsdb} and \texttt{manticore}.
An explanation for this is that, as shown in Table \ref{tab:function_analysis}, the former 
repositories have significantly higher numbers of appropriate functions than the latter. 
This might introduce more variability into the results for the former repositories. 
Code coverage is also rather good. It ranges from about 68\% to greater than 80\%. 
Importantly, this code coverage is obtained without knowledge of the bodies of these functions, only their type signatures.

\renewcommand{\arraystretch}{1}
\begin{table}[htb]
\small
\centering
\begin{tabular}{lccc}
\toprule
\textbf{Repository} & \textbf{Total Functions} & \textbf{Annotated Functions} & \textbf{Appropriate Functions} \\
\midrule

mypy       & 1028 & 1028 & 132 \\
mindsdb    & 400  & 55   & 5   \\
black      & 248  & 248  & 35  \\
manticore  & 211 & 26   & 5   \\
\bottomrule
\end{tabular}
\caption{Function breakdown by repository.}
\label{tab:function_analysis}
\end{table}

Table \ref{tab:function_analysis} presents the breakdown of function totals across the four repositories studied. Importantly, a "function" in this context is a top-level function defined in Python. We do not currently consider functions that are defined as methods of classes. These are not counted in the totals, and they are not eligible to be appropriate functions. Python also supports the definition of nested functions. Functions that are nested within other functions are not counted in the total number of functions and are not eligible to be appropriate functions.

Although we obtained good coverage overall, there were certain function bodies for which we achieved lower coverage. Broadly, the reasons for lower coverage that we have observed can be bucketed into 1) overly broad type annotations that do not represent the data the function is expecting, 2) external dependencies that the function has on either the file system or global program state, and 3) complex branch conditions that are hard to satisfy. 

\begin{figure}[htb]
\centering
\includegraphics[width=4in]{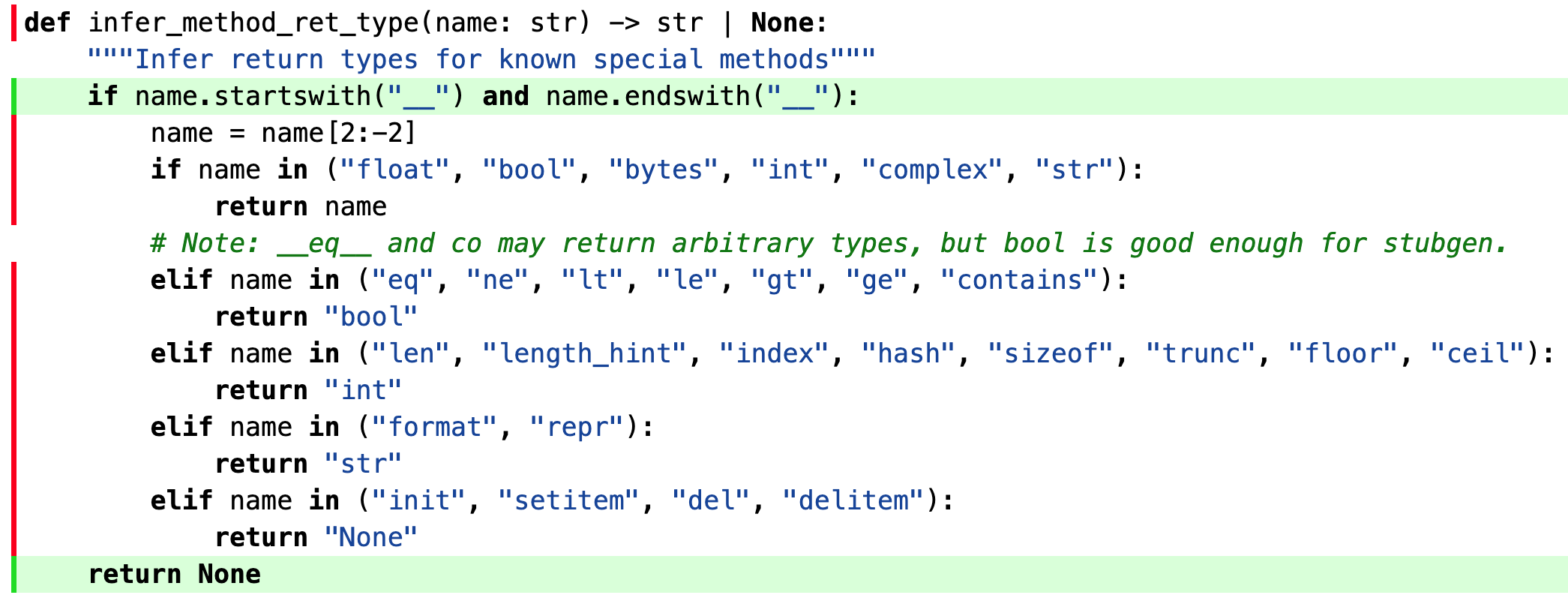}
\caption{Complex string condition that is difficult to pass when the input can be any string.}\label{fig:mypycovexample}
\end{figure}

Figure \ref{fig:mypycovexample} contains the \textit{coverage.py} report output for one of the appropriate functions in \texttt{mypy} that was fuzzed, \texttt{infer\_method\_ret\_type}. The low coverage here is because of the complex branch condition, which is checking whether the given string starts and ends with two underscore characters.

\begin{figure}[htb]
\centering
\includegraphics[width=4in]{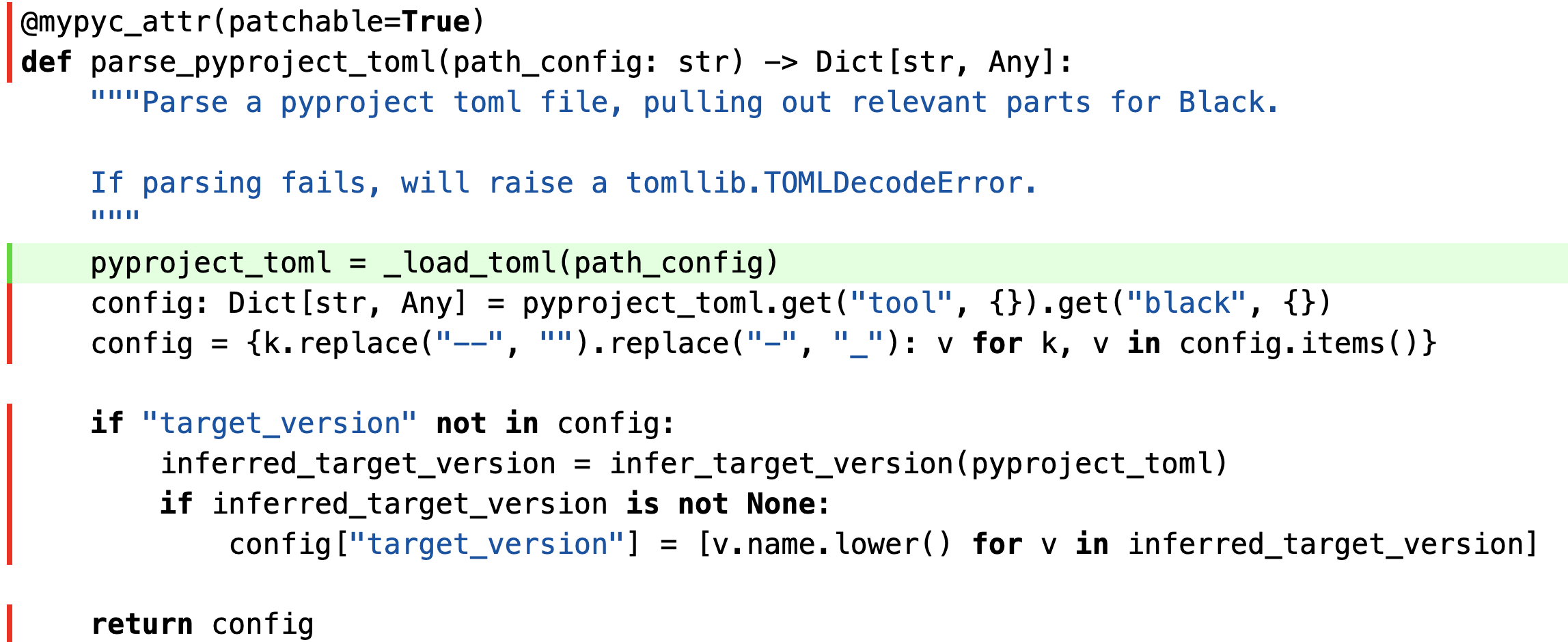}
\caption{Unmet file system dependency causing function call to fail.}\label{fig:blackcovexample}
\end{figure}

Figure \ref{fig:blackcovexample} shows an example of low coverage from the \texttt{black} repository which is not getting fully covered due to an unmet file system dependency. In this case, the argument to the function, \texttt{path\_config}, represents a path to a valid TOML file. It is highly unlikely to spontaneously generate a valid file path, and we do not intentionally set up TOML files in a test bed for fuzzing. Therefore, execution results in an exception and the remainder of the function is not covered.

\begin{figure}[htb]
\centering
\begin{subfigure}[b]{0.45\textwidth}
    \centering
    \includegraphics[width=\textwidth]{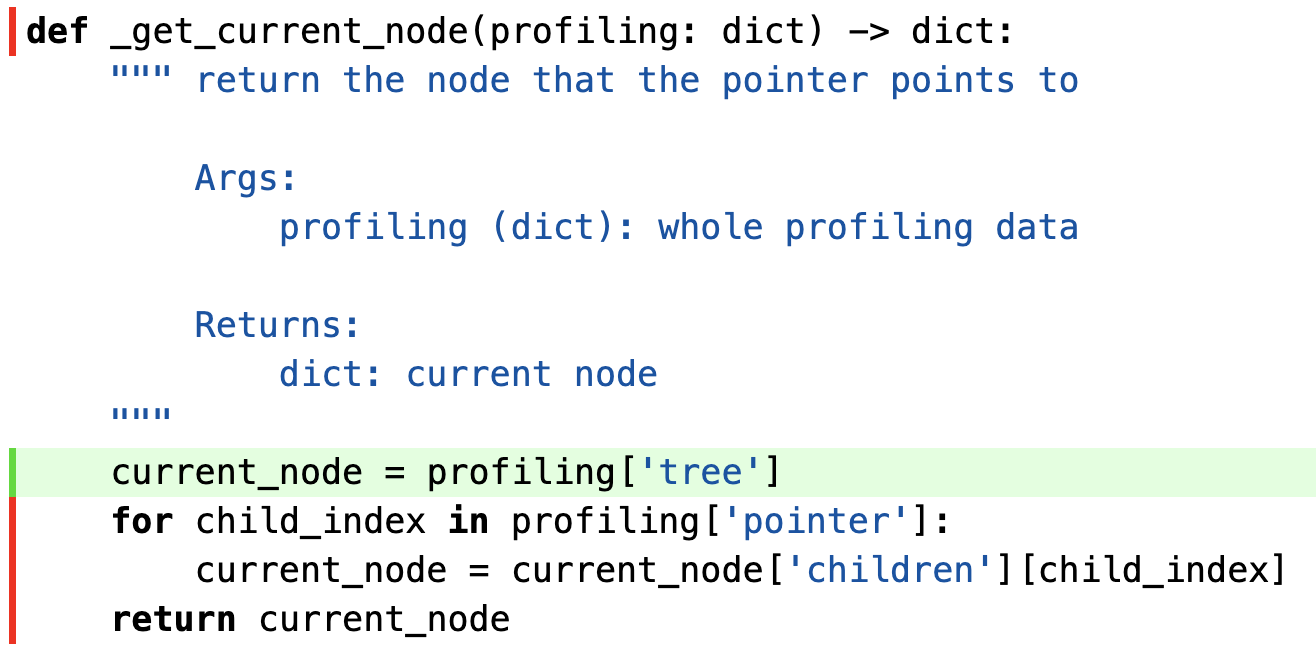}
    \caption{Dictionary lookup failed due to implicit dictionary structure.}
    \label{fig:mindsdbcovexample}
\end{subfigure}
\hfill
\begin{subfigure}[b]{0.45\textwidth}
    \centering
    \includegraphics[width=\textwidth]{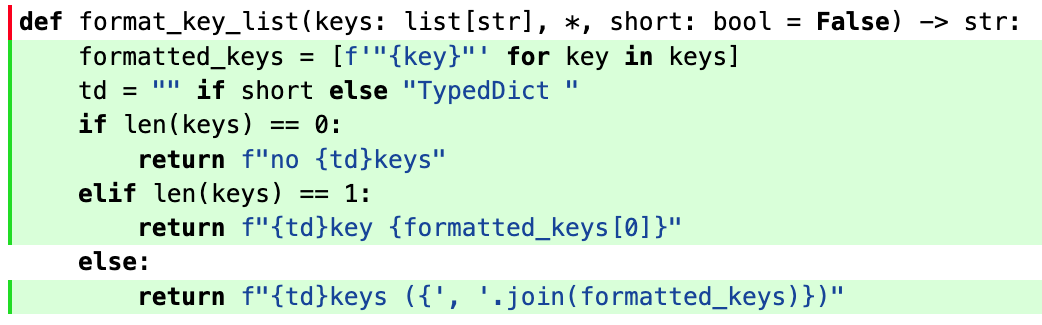}
    \caption{Function with no external dependencies and simple conditionals is fully covered.}
    \label{fig:mypyeasycovexample}
\end{subfigure}
\caption{Additional code coverage examples.}
\end{figure}

Figure \ref{fig:mindsdbcovexample} contains an example of low code coverage from \texttt{mindsdb}. This example represents a deficiency in the parameter type annotation of the function. The function expects that \texttt{'tree'} and \texttt{'pointer'} are both present as keys in the given dictionary, but the type annotation broadly specifies a dictionary with keys and values of any type. A \texttt{KeyError} exception is thrown that interrupts execution.

Finally, Figure \ref{fig:mypyeasycovexample} shows an example where we obtain full coverage of the function body. In this
case, the enumerator for the \texttt{keys} argument, which is a list of strings, produces values that cover the
three main branches in the function: keys having length 0, 1, or more than 1.

\section{Discussion and Future Work}\label{sect:future_work}
\begin{codefloat}
\begin{code}[language=Python]
def create_decimal(whole_part: int, decimal_part: float) -> float:
    """Create a decimal number from the given whole part and decimal part"""
    return float(str(whole_part) + str(decimal_part).lstrip('0'))

def test_create_decimal_no_exception(x: int, y: float) -> bool:
    """Property-based test to ensure create_decimal doesn't throw (obviously, this fails)"""
    try: 
        create_decimal(x, y)
    except: 
        return False
    return True
\end{code}
\caption{Example of a property-based test that could be serviced by the enumerative embedding.}\label{listing:propertybasedtest}
\end{codefloat}

In this paper, we introduce an enumerative embedding of the Python type system in ACL2s.
 In our estimation, the principal application of the embedding, because it is enumerative, is fuzzing. 
Fuzzing requires high-quality inputs to be successful, and the customizability of enumerators 
enables users to create representative examples of their data \cite{xifarasleveraging2024}. 

In light of its use as the foundation for fuzzing, we validated in our evaluation (Section \ref{sect:evaluation}) that the inputs generated by our custom enumerators cover code effectively in functions whose type signatures are embedded. However, we also found that code coverage is limited by type annotations that are too broad, external dependencies on program or system state, and complex branch conditions. These can be addressed in future work in the following ways:
\begin{enumerate}
\item \textbf{Overly broad annotations:} Analyze the errors that are raised when sending inputs into functions, or send in mock objects that are instrumented to track how they are used, to constrain the type definitions.
\item \textbf{External dependencies:} Implement mocks of global program state and the file system that the code being tested can interact with.
\item \textbf{Complex branch conditions:} Extract branch conditions, embed them in ACL2s, and use ACL2s to produce examples that satisfy and do not satisfy them.
\end{enumerate}

A threat to the validity of these results is that the latest source code for the custom enumerators 
is different from the enumerators that were used to collect this data, but the adjustments are 
minor enough that we do not anticipate significant effects on the results. 

The validity of the results and the embedding itself are also limited by the set of supported types. 
A core set of types has been implemented, but there are many types in Python's 
\texttt{typing} module, such as \texttt{Sequence} and \texttt{Iterable}, that are used often in 
Python code. In particular, function types are important because functions are 
first-class objects in Python. They are denoted in type annotations using the \texttt{Callable} 
annotation in the \texttt{typing} module. This embedding becomes significantly more usable on
the average repository when these types are embedded. This is a top priority for future work.

Given its suitability for fuzzing, this embedding further enables property-based testing for typed properties written as 
Python functions. For instance, Listing \ref{listing:propertybasedtest} specifies the property 
"\texttt{create\_decimal()} does not throw an exception." Examples for Python integers and 
floats generated by enumerators can be streamed as inputs to \texttt{test\_create\_decimal\_no\_exception}, and if 
this function returns False, the property is violated. This testing methodology represents a practical compromise
between unit testing, where fixed scenarios with strong assumptions are tested, and formal verification.
Property-based testing libraries exist for Python~\cite{maciverhypothesis2019, hatfield-doddshypofuzz2022}, and 
we look forward to evaluating opportunities for integration and collaboration in future work. 

Another compelling application for this embedding is type checking. Once the full type system semantics 
are embedded and additional information about the code such as basic control flow skeletons is extracted, 
a theorem-prover-based type checker could be built which may have better soundness and completeness 
properties than current solutions. We leave this as another exciting direction for future work. 

\section{Conclusion}\label{sect:conclusion}
As Python continues to grow in popularity, the ability to test applications written in it grows in importance. Meanwhile, the growing prevalence of developer-added type annotations in Python renders automated analyses more tractable~\cite{digraziaevolution2022}. 
In this paper, we enable tool developers to leverage this situation with an embedding of the Python type system in ACL2s. 
This embedding is enumerative, meaning that examples of types can be generated easily. 
This enables dynamic testing of Python code, which is useful in the absence of a formal model of 
Python's complex semantics. The embedding is additionally extensible. We invite the community 
to extend this embedding with additional types and typing constructs. Python's documentation contains 
information on what types and typing constructs are available in the language~\cite{pythonsoftwarefoundationtypingnodate}. 

Our long-term goal is to create data definitions and enumerators that support the entire type system. 
Eventually, this embedding may serve as a foundation for advanced property-based testing and 
reasoning in Python, and we hope to advance further toward this vision in future work. 

\section*{Acknowledgments}

We would like to acknowledge the MIT SuperCloud team for granting us access to their high-performance 
computing environment on which we ran our experiments \cite{reuther2018interactive}.

\bibliographystyle{eptcs}
\bibliography{bibliography-cleaned,standard-entries}
\end{document}